\documentclass[runningheads]{llncs}
\usepackage[T1]{fontenc}
\usepackage{graphicx}
\usepackage{amsmath}
\usepackage{multirow}
\usepackage[hidelinks]{hyperref}
\usepackage{color}

\begin{document}
\title{Improving the U-Net Configuration for Automated Delineation of Head and Neck Cancer on MRI}
\titlerunning{Improving the U-Net Configuration for MRI Segmentation}
\author{Andrei Iantsen\inst{1}\orcidID{0000-0002-6690-6070} 
}
\authorrunning{A. Iantsen}
\institute{Independent Researcher\\
\email{andrei.iantsen@gmail.com}\\
}
\maketitle              % typeset the header of the contribution
\begin{abstract}
Tumor volume segmentation on MRI is a challenging and time-consuming process that is performed manually in typical clinical settings. This work presents an approach to automated delineation of head and neck tumors on MRI scans, developed in the context of the MICCAI Head and Neck Tumor Segmentation for MR-Guided Applications (HNTS-MRG) 2024 Challenge. Rather than designing a new, task-specific convolutional neural network, the focus of this research was to propose improvements to the configuration commonly used in medical segmentation tasks, relying solely on the traditional U-Net architecture. The empirical results presented in this article suggest the superiority of patch-wise normalization used for both training and sliding window inference. They also indicate that the performance of segmentation models can be enhanced by applying a scheduled data augmentation policy during training. Finally, it is shown that a small improvement in quality can be achieved by using Gaussian weighting to combine predictions for individual patches during sliding window inference. The model with the best configuration obtained an aggregated Dice Similarity Coefficient (DSCagg) of 0.749 in Task~1 and 0.710 in Task~2 on five cross-validation folds. The ensemble of five models (one best model per validation fold) showed consistent results on a private test set of 50 patients with an DSCagg of 0.752 in Task~1 and 0.718 in Task~2 (team name: {\textsf{andrei.iantsen}}). The source code and model weights are freely available at \url{www.github.com/iantsen/hntsmrg}.

\end{abstract}
\section{Introduction}
Radiation therapy (RT) plays a crucial role in oncology with more than 40\% of patients worldwide undergoing RT at least once as part of cancer treatment~\cite{Thompson2018}.  Modern linear accelerators can deliver radiation beams to tissues with submillimeter precision and further advances in RT necessitate the integration of increasingly accurate imaging systems for tumor targeting. Computed tomography (CT) is widely used for cancer staging and RT planning. However, due to the limited contrast between soft tissues on CT, magnetic resonance imaging (MRI) is often applied instead of or in addition to CT to better distinguish the tumor from surrounding normal tissues in anatomical areas, such as the brain, nasopharynx and pelvis. Moreover, emerging MR-guided linear accelerators can monitor the target volume and organs at risk in real-time during dose delivery and adjust the treatment plan daily. Despite the good contrast and high spatial resolution, tumor volume segmentation on MRI scans is a challenging and time-consuming process that is performed manually in typical clinical settings. Consequently, the resulting tumor contours are subject to significant intra- and inter-observer variability, which can lead to deleterious consequences in downstream applications (e.g., skewed dose distributions during RT planning; low repeatability/reproducibility of image-based biomarkers in radiomics~\cite{Traverso2018,Desseroit2016}). Hence, fully automated methods for MRI segmentation are of particular interest from a clinical perspective.

Due to the rapid advances in deep learning and computing technologies over the last decade, data-driven models based on convolutional neural networks (CNNs) have achieved impressive results in a wide range of computer vision tasks, including image segmentation. In the medical imaging domain, U-Net has remained a workhorse since its introduction in 2015~\cite{unet}. Furthermore, despite a variety of alternative, task-specific models reported in the literature, the vast majority of them actually constitute some variants of U-Net, often with only cosmetic changes. Finally, as shown in the nnU-Net framework~\cite{nnunet_nature,Isensee2024}, other components of the overall configuration (e.g., data pre- and post-processing methods, augmentation techniques, training procedures, etc.) often have a greater impact on performance than the choice of architecture per se.

This paper presents an approach to automated delineation of head and neck cancer on MRI, developed in the context of the MICCAI Head and Neck Tumor Segmentation for MR-Guided Applications (HNTS-MRG) 2024 Challenge. The main goal of this research was to propose some improvements to the configuration commonly used in medical segmentation tasks, relying only on the traditional U-Net architecture without significant changes.

\section{HNTS-MRG 2024 Challenge}
\subsection{Data Description} For the purpose of the challenge, the University of Texas MD Anderson Cancer Center provided a dataset of 150 patients with histologically proven head and neck cancer~\cite{dataset}. Two T2-weighted (T2w) MRI scans were available for each patient: a pre-RT scan acquired 1--3 weeks before RT and a mid-RT scan after 2--4 weeks of RT. Manual delineation of primary gross tumor volume (GTVp) and metastatic lymph nodes (GTVn) for each patient was performed independently by 3 to 4 clinical experts, whose results were then combined using  the STAPLE algorithm~\cite{staple}. The resulting segmentation with three target classes (background, GTVp, GTVn) served as the ground truth. 
\subsection{Segmentation Tasks} Two segmentation tasks were proposed in the HNTS-MRG 2024 Challenge. In \textbf{Task~1}, it was required to build an automated solution for segmenting GTVp and GTVn volumes only on pre-RT scans. While in \textbf{Task~2}, the goal was to delineate the target volumes on mid-RT scans, optionally using the corresponding pre-RT scans and ground truth annotations for them as input data. 
\subsection{Evaluation Metric}  The Dice Similarity Coefficient (DSC) is a widely used metric for evaluating performance in segmentation tasks. For a binary ground truth $y$ and a binary prediction $\hat{y}$, the DSC is calculated as
\begin{equation}
\text{DSC} (y, \hat{y})= 
2\frac{\sum\limits_{i} y_{i}\hat{y}_{i}}{\sum\limits_{i} y_{i} + \sum\limits_{i} \hat{y}_{i}}\,,
\end{equation} 	
where $y_{i}$ and $\hat{y}_{i}$ are the true and predicted labels for the $i$th element (voxel), respectively. If the ground truth has no elements of the target class (i.e., $y_{i} = 0$ for any $i$), this metric is not informative since $\text{DSC} = 0$ for any prediction $\hat{y}$. Accordingly, the aggregated Dice Similarity Coefficient ($\text{DSC}_{\text{agg}}$) was used for evaluation in the challenge. For a set of $N_{\mathcal{S}}$ pairs of binary ground truth and predicted masks, $\mathcal{S} = \{(y^{(n)}, \hat{y}^{(n)})\}_{n=1}^{N_{\mathcal{S}}}$, the $\text{DSC}_{\text{agg}}$ is defined as
\begin{equation}
	\text{DSC}_{\text{agg}} (\mathcal{S})= 
	2\frac{\sum\limits_{n, i}  y^{(n)}_{i}\hat{y}^{(n)}_{i}}{\sum\limits_{n, i}y^{(n)}_{i} + \sum\limits_{n, i}\hat{y}^{(n)}_{i}}\,.
\end{equation} 	
The average $\text{DSC}_{\text{agg}}$ for the GTVp and GTVn classes on a test set of 50 patients was used to evaluate performance in both tasks presented in the challenge. 

\section{Methods}
\subsection{Network Architecture}
\begin{figure}
	\includegraphics[width=\textwidth]{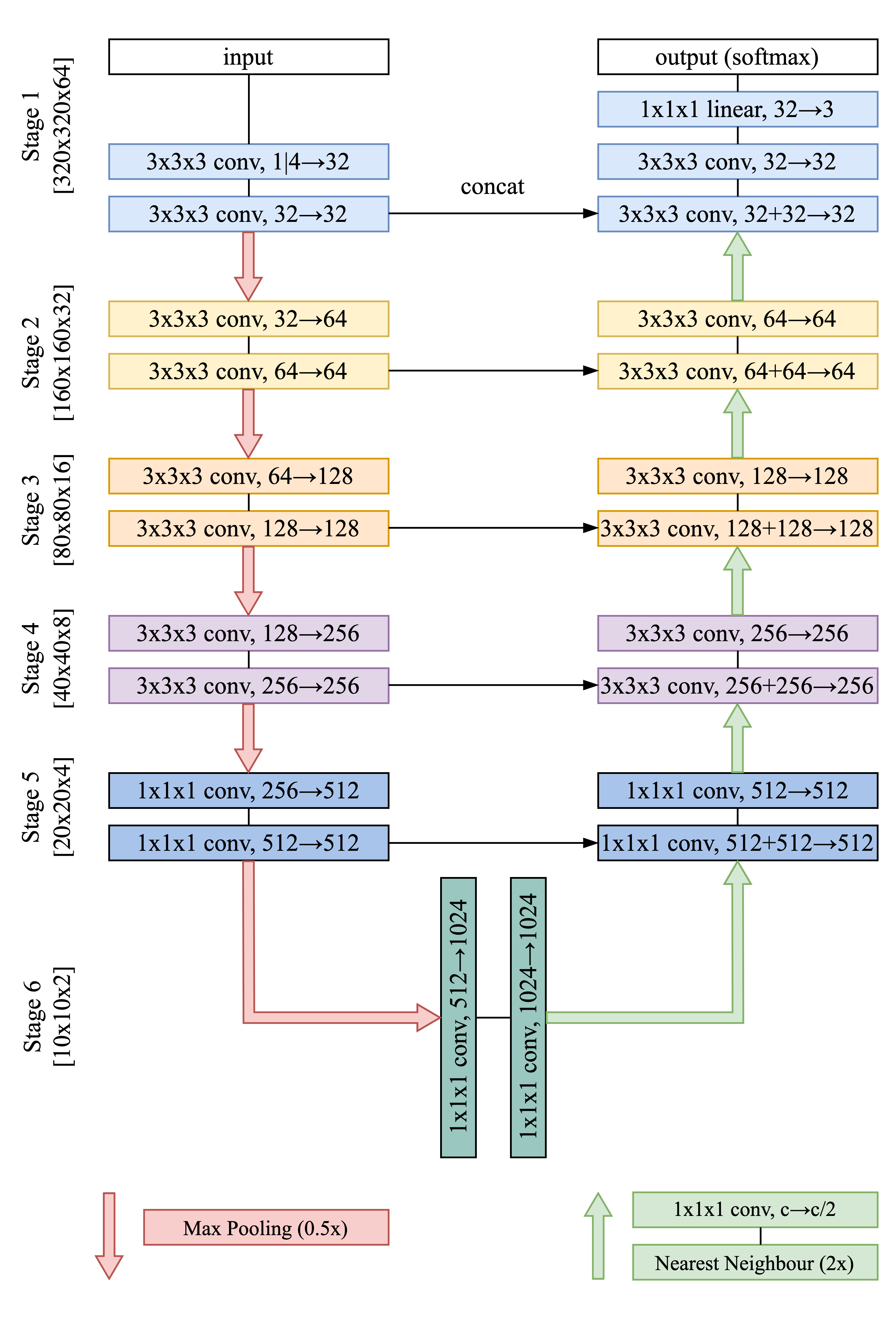}
	\caption{The network architecture used in both task. It is the traditional U-Net, in which convolutional blocks in Stages 5--6  are implemented with $1\times1\times1$ kernels to reduce the number of model parameters. Spacial sizes of feature maps are provided in square brackets. } \label{network}
\end{figure}

The network used for both tasks followed the design principles of the traditional U-Net~\cite{unet} (see
Fig.~\ref{network}). It was built using convolutional blocks, each consisting of a 3D convolutional layer, instance normalization, and ReLU nonlinearity. In the encoder, the number of feature maps (i.e., channels) was doubled after downsampling, which was preformed with a $2\times2\times2$ max pooling. Upsampling in the decoder was carried out with a $1\times1\times1$ convolutional block to halve the number of feature maps, followed by nearest-neighbor interpolation to double their spatial size (resolution). Feature maps from five resolution stages in the encoder were transferred to the decoder via skip-connections. Convolutional blocks in Stages 1--4 were implemented with $3\times3\times3$ kernels, whereas smaller kernels of size $1\times1\times1$ were employed in Stages 5--6 to substantially decrease the number of model parameters (86M to 14M). The softmax activation was applied to generate probability scores for three output classes.

\subsection{Cross-Validation Folds}
The provided dataset was divided into five equal folds for cross-validation, with all images for each individual patient placed in only one fold. The results on the validation folds were used to compare different configurations and evaluate the generalization performance of the model (i.e., the expected performance on new data examples).

\subsection{Training}
Before being processed by the model, all the MRI scans and segmentation masks were first resampled to a voxel size of $0.5\times0.5\times2$ mm using linear and nearest-neighbor interpolation methods, respectively. Model training was done on patches of size $320\times320\times64$ voxels that were randomly sampled from the entire images. The position of each patch was chosen such that 90\% of training patches contained some voxels of a target class (i.e., GTVp or GTVn), while the remaining 10\% were extracted completely randomly.  

For Task~1, the network was trained on all the provided MRI scans (i.e., pre-RT, mid-RT, and pre-RT registered to mid-RT) for 100K iterations (batches) with a batch size of 2. The model for Task~2 was trained for 50K iterations and had four input channels: a mid-RT scan, a registered pre-RT scan, and two binary masks for GTVp and GTVn on the registered pre-RT. Performance on validation examples was evaluated after every 5K training iterations. For both tasks, Adam optimizer~\cite{adam} was used with a learning rate decreasing from $10^{-3}$ to $10^{-5}$ following the cosine decay schedule~\cite{cosine}. The model training was performed on a single GPU with 16 GB of VRAM using mixed precision~\cite{mixedprecision} to significantly reduce the required memory and shorten the execution time.

\subsection{Loss Function}
During training, the loss function was the Dice Loss computed on the entire batch of ground truth masks and model predictions for each class, $\mathcal{B} = \{(y^{(n)}, \hat{p}^{(n)})\}_{n=1}^{N_{\mathcal{B}}}$:
\begin{equation}\label{loss}
	{L} _{Dice}(\mathcal{B})= 1 - 2\frac{\sum\limits_{n, i} y^{(n)}_{i}\hat{p}^{(n)}_{i}}{\sum\limits_{n, i} y^{(n)}_{i} + \sum\limits_{n, i} \hat{p}^{(n)}_{i}}\,.
\end{equation}

The second term in Eq.~\ref{loss} is the smooth approximation of the $\text{DSC}_{\text{agg}}$  function for one class in batch $\mathcal{B}$. Because data examples can have only a subset of classes in their ground truth masks (e.g., one or both tumor volumes may be completely missing in some patients, or the tumor may not be in the patch due to sampling), the average loss is calculated only for classes present in the training batch, including the background.

\subsection{Sliding Window Inference} 
Inference on entire MRI scans was performed relying on a sliding window approach: predictions were obtained for consecutive patches of size $320\times320\times64$ using a stride of $80\times80\times16$ voxels. Note that predictions on overlapping voxels can be combined in different ways. The default option is to average them by assigning equal weights to all voxels. However, it is known that the accuracy of patch-based predictions decreases towards the patch edges, which can lead to different artifacts on the combined output mask~\cite{edgeart}. Alternative methods are based on weighting voxels according to their positions in the patch, so that voxels closer to the patch center have higher weights. For both tasks, predictions for individual patches were combined using the \textit{Gaussian weights} ranging from 1 at the patch center to 0.1 at the edges. This approach slightly improved the equality of sliding window predictions compared to equal weighting (see Section~\ref{results}). After inference, the model predictions were converted into class labels by applying the argmax function.

\begin{figure}
	\includegraphics[width=\textwidth]{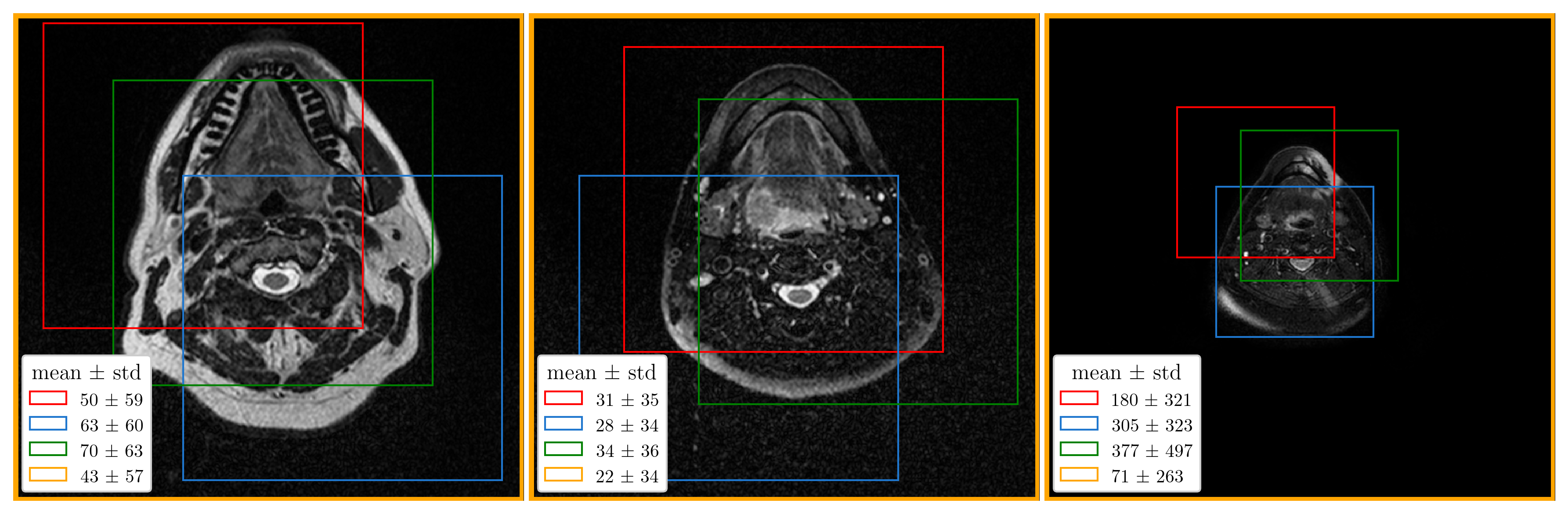}
	\caption{The mean and standard deviation computed for different patches (first three contours), each of size $320\times320\times64$,  and entire images (last contours) from the training set after resampling to the same voxel size. An axial slice is shown for each example.} \label{patches}
\end{figure}

\subsection{Intensity Normalization}
In contrast to CT, the intensity scale in MRI is not standardized and therefore intensity normalization is particularly important, especially when working with images from different MRI scanners~\cite{Wahid2021}. Furthermore, input normalization is generally required to improve the convergence of optimization methods based on gradient descent. Z-score normalization, where all intensities are first shifted by the mean and then scaled by the standard deviation, works well in practice. However, this type of normalization is often applied with the mean and standard deviation computed on the entire image (\textit{image-wise normalization}), even when training is carried out on image patches (see nnU-Net~\cite{nnunet_nature}). As a result, the intensity distribution in each patch (i.e., its mean and variance) depends on the patch location, which can significantly hinder the convergence of optimization methods. Similarly, the performance of models trained with \textit{patch-wise normalization} (i.e., normalization is done \textit{after} patch extraction) is affected by covariate shift~\cite{covshift,covshift2} when inference is performed on normalized images in a sliding window manner. Fig.~\ref{patches}  shows the differences in means and standard deviations between different patches and between images from the training set. To mitigate the effect of covariate shift, only patch-wise normalization was applied to the model inputs in both tasks (except for the binary masks in Task~2). At inference, the normalization was integrated directly into the sliding window approach.

\subsection{Data Augmentation}
Data augmentation was used to increase the size and diversity of training examples. Transformations such as image mirroring, rotations, contrast adjustment, imitation of MRI bias field and motion artifacts, as well as distortions by additive Gaussian noise were applied. Each transformation was performed independently with the same probability, which was linearly increasing from 0.05 to 0.25 over the course of training with adjustments made every 1K batches. This \textit{scheduled augmentation policy} was aimed to generate more data examples towards the end of training when overfitting was more likely to occur. The comparison between the scheduled data augmentation and the augmentation with a constant probability of 0.15 is provided in Section~\ref{results}.

\subsection{Model Ensembling}
The configuration with the best performance on the validation folds was used for the final submission in the challenge. Predictions were obtained by averaging the softmax outputs of five models (one best model per fold) and applying the argmax function to obtain the class labels. The output was then resampled using nearest-neighbor interpolation to restore the original resolution (voxel size) of the input. The source code and model weights to reproduce the submitted results are freely available at \url{www.github.com/iantsen/hntsmrg}.

\section{Results and Discussion}\label{results}
The main focus of this research, largely inspired by the nnU-Net framework, was to improve some components of the configuration commonly used in medical segmentation tasks without making significant changes to the architecture of the traditional U-Net. The following configuration was used as a baseline for comparison: (1) the intensities of MRI scans were normalized image-wise (i.e., before patch extraction), (2) data augmentation was applied with a probability of 0.15, which remained constant during training, and (3) all voxels in the patch had equal weights during sliding window inference.

The empirical results for different configurations on five validation folds are summarized in Table~\ref{tab_val}. The use of patch-wise normalization for both training and inference improved performance in both tasks compared to the baseline: the average $\text{DSC}_{\text{agg}}$ increased from 0.689 to 0.725 in Task~1, and from 0.698 to 0.712  in Task~2. This improvement can be attributed to a reduction in covariate shift, as this type of normalization guarantees that all inputs have zero mean and unit variance. 
Training the model on a progressively increasing number of examples created with data augmentation techniques produced better results in Task~1 (the metric changed from 0.725 to 0.745), but had no significant impact in Task~2 (0.712 vs. 0.711). Similarly, using the Gaussian weights to combine predictions for individual patches during sliding window inference led to slightly higher results in Task~1 (0.745 vs. 0.749), but not in Task~2 (0.711 vs. 0.710). As for the results for each target class, the configuration with all three modifications achieved more accurate predictions for GTVn than for GTVp in Task~1 (0.804 and 0.695, respectively) and in Task~2 (0.823 and 0.598, respectively). Surprisingly, the model showed significantly worse results for GTVp in Task~2, although registered pre-RT masks were provided as inputs, compared to Task~1 where segmentation was performed only on pre-RT scans. The configuration with all three modifications was selected for the final submission in both tasks.

\begin{figure}
	\includegraphics[width=\textwidth]{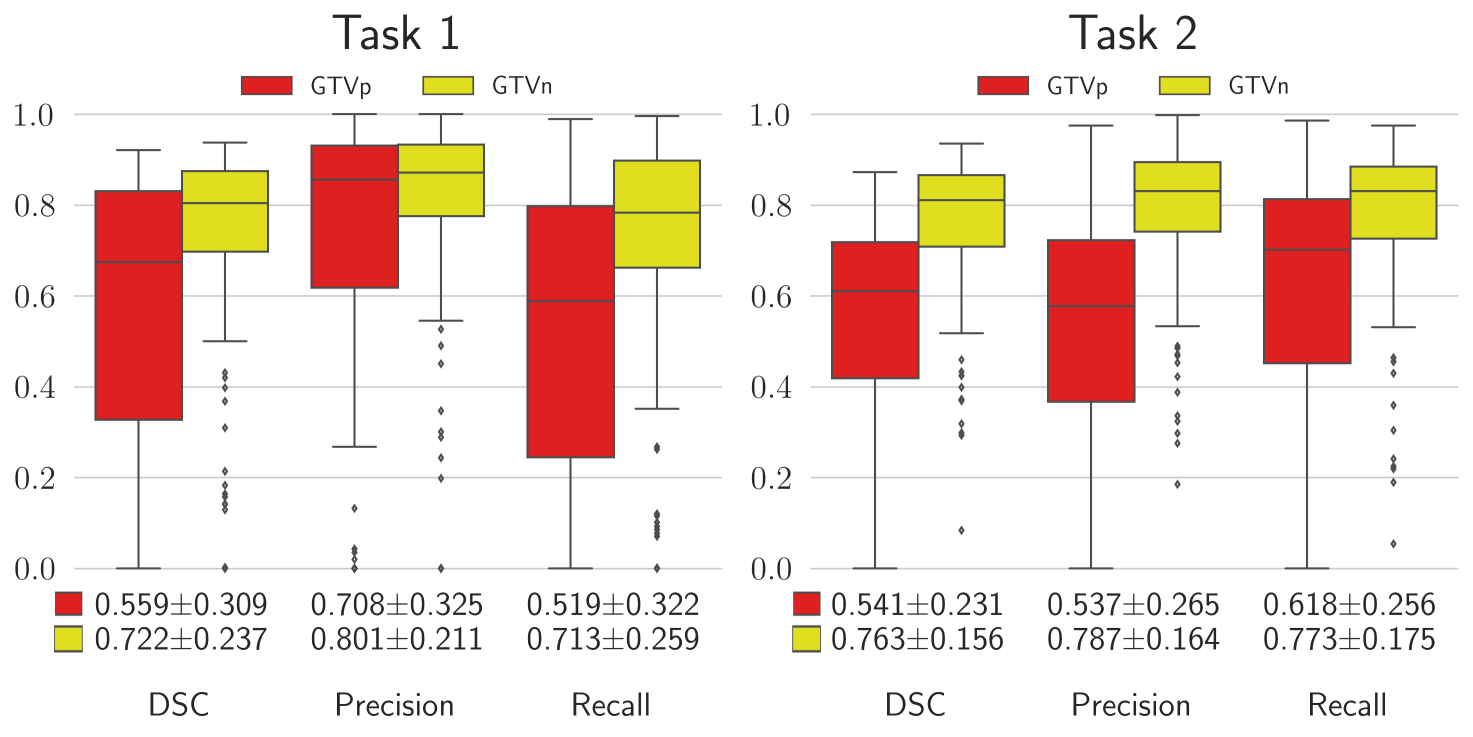}
	\caption{The distribution of results for patients from five validation folds. All metrics were calculated using only examples with non-zero ground truth masks.} \label{boxplot}
\end{figure}

\begin{figure}
	\includegraphics[width=\textwidth]{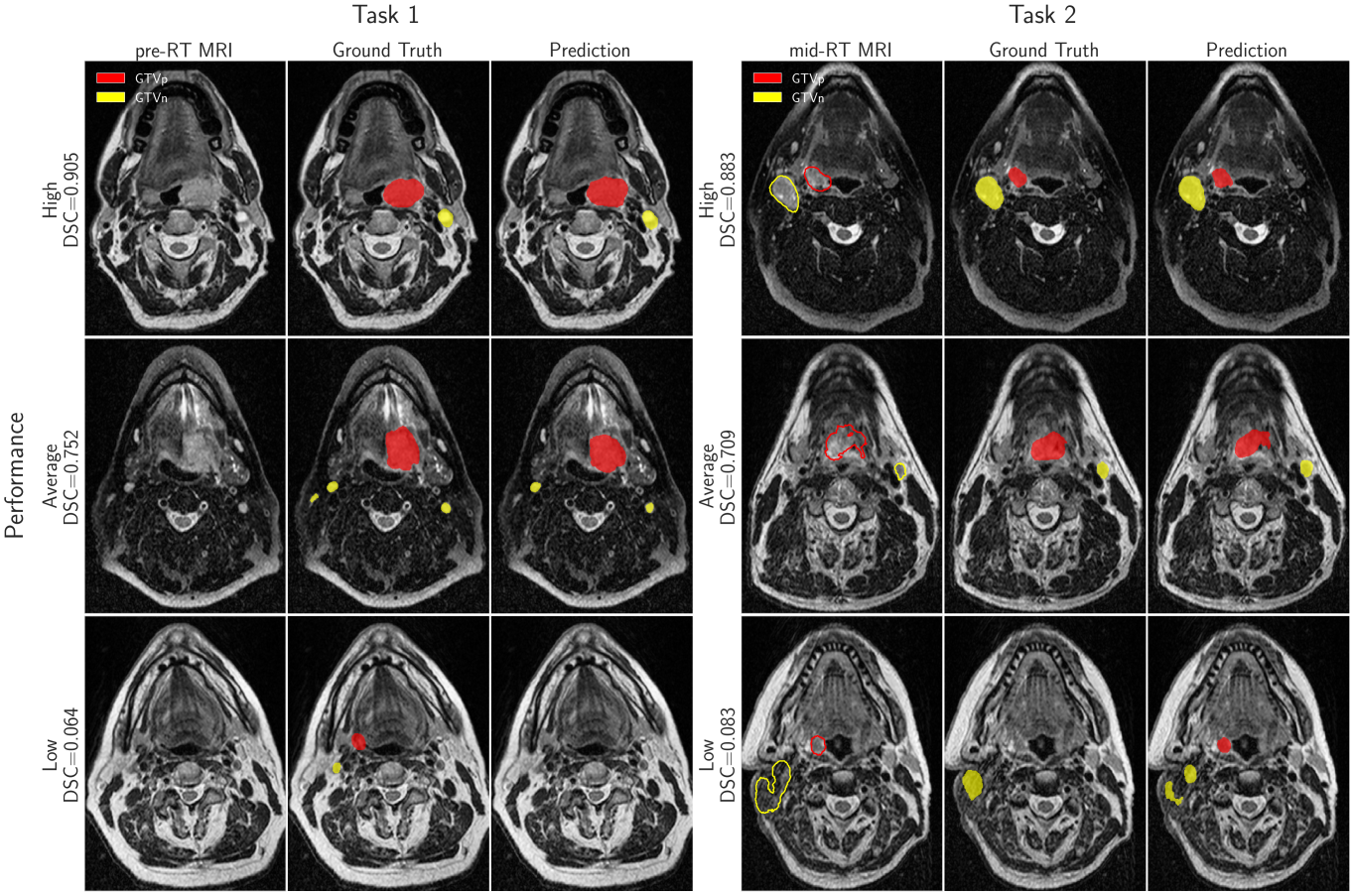}
	\caption{Examples for visual comparison of model predictions and ground truth mask for both tasks (filled contours). ``DSC'' is the average DSC for the GTVp and GTVn classes. Contours of both target classes from registered pre-RT scans are drawn on corresponding mid-RT scans (Task~2, unfilled contours). Examples of ``average performance'' were selected based on the average $\text{DSC}_{\text{agg}}$ on validation folds in both tasks.} \label{visual}
\end{figure}

In addition to the aggregated metrics in Table~\ref{tab_val}, Fig.~\ref{boxplot} shows the results in terms of DSC, Precision, and Recall for individual data examples (i.e., patients) from the validation folds. A visual comparison of the predicted and ground truth masks for three different patients is provided in Fig.~\ref{visual}.

The final results on the test set of 50 patients are shown in Table~\ref{tab_test}. The ensemble of five models achieved an average $\text{DSC}_{\text{agg}}$ of 0.752 in Task~1 (0.709 and 0.794 for the GTVp and GTVn classes, respectively) and 0.718 in Task~2 (0.592 and 0.845).

\begin{table}[t]
	\centering
	\caption{Average values of $\text{DSC}_{\text{agg}}$ for the GTVp and GTVn classes on \textit{five validation folds}. ``Baseline configuration'' refers to training with the image-wise normalization of input intensities and the augmentation policy with a constant probability of 15\%, as well as with equal weights for all voxels in the patch during sliding window inference.}\label{tab_val}
	\begin{tabular}{l|c c c|c c c}
		\hline
		\multirow{2}{*}{\textbf{Configuration}} & \multicolumn{3}{|c|}{\textbf{Task~1}} & \multicolumn{3}{c}{\textbf{Task~2}}\\ 
		&  \textbf{GTVp} & \textbf{GTVn} & \textbf{Average} &  \textbf{GTVp} & \textbf{GTVn} & \textbf{Average} \\
		\hline
		Baseline & 0.672 & 0.705 & 0.689 & 0.581 & 0.815& 0.698 \\
		\hline
		+ patch-wise normalization &  0.682 & 0.768 & 0.725 & 0.600 & 0.824 & 0.712 \\
		+ scheduled augmentation  &  0.691 & 0.800 & 0.745 & 0.598 & 0.825 & 0.711 \\
		+ Gaussian weighting &  0.695 & 0.804 & 0.749 & 0.598 & 0.823 & 0.710 \\
		\hline
	\end{tabular}
\end{table}

\begin{table}[t]
	\centering
	\caption{Average values of $\text{DSC}_{\text{agg}}$ for the GTVp and GTVn classes on the \textit{test set} with 50 patients. Predictions for each task were obtained using the ensemble of five models.}\label{tab_test}
	\begin{tabular}{c c c|c c c}
		\hline
		\multicolumn{3}{c|}{\textbf{Task~1}} & \multicolumn{3}{c}{\textbf{Task~2}}\\ 
		\textbf{GTVp} & \textbf{GTVn} & \textbf{Average} &  \textbf{GTVp} & \textbf{GTVn} & \textbf{Average} \\
		\hline
		0.709 & 0.794 & 0.752 & 0.592 & 0.845 & 0.718\\
		\hline
	\end{tabular}
\end{table}

\section{Conclusion}
This paper presents a number of modifications to a configuration commonly used with convolutional neural networks for medical image segmentation. The empirical results were obtained using the traditional U-Net architecture to address two segmentation tasks proposed in the context of the MICCAI Head and Neck Tumor Segmentation for MR-Guided Applications (HNTS-MRG) 2024 challenge.  First, it was shown that patch-wise normalization (i.e., normalization applied after patch extraction) used for both training and sliding window inference improved performance by reducing covariance shift. Second, the scheduled data augmentation policy, where each transformation was applied with a probability linearly increasing towards the end of training, produced better results compared to augmentation with a fixed probability. Finally, using Gaussian weighting to combine predictions for individual patches during sliding window inference resulted in slightly more accurate predictions than those obtained with equal weighting.

\bibliographystyle{splncs04}
\bibliography{paper}

\begin{thebibliography}{10}
\providecommand{\url}[1]{\texttt{#1}}
\providecommand{\urlprefix}{URL }
\providecommand{\doi}[1]{https://doi.org/#1}

\bibitem{Desseroit2016}
Desseroit, M.C., Tixier, F., Weber, W.A., Siegel, B.A., Cheze Le~Rest, C.,
  Visvikis, D., Hatt, M.: Reliability of {PET/CT} shape and heterogeneity
  features in functional and morphologic components of non–small cell lung
  cancer tumors: A repeatability analysis in a prospective multicenter cohort.
  Journal of Nuclear Medicine  \textbf{58}(3),  406–411 (Oct 2016).
  \doi{10.2967/jnumed.116.180919},
  \url{http://dx.doi.org/10.2967/jnumed.116.180919}

\bibitem{nnunet_nature}
Isensee, F., Jaeger, P.F., Kohl, S.A.A., Petersen, J., Maier-Hein, K.H.:
  {nnU-Net}: a self-configuring method for deep learning-based biomedical image
  segmentation. Nature Methods  \textbf{18}(2),  203–211 (Dec 2020).
  \doi{10.1038/s41592-020-01008-z},
  \url{http://dx.doi.org/10.1038/s41592-020-01008-z}

\bibitem{Isensee2024}
Isensee, F., Wald, T., Ulrich, C., Baumgartner, M., Roy, S., Maier-Hein, K.,
  J\"{a}ger, P.F.: {nnU-Net} revisited: A call for rigorous validation in 3d
  medical image segmentation. In: Medical Image Computing and Computer Assisted
  Intervention – {MICCAI} 2024. p. 488–498. Springer Nature Switzerland
  (2024). \doi{10.1007/978-3-031-72114-4_47}

\bibitem{adam}
Kingma, D.P., Ba, J.: Adam: A method for stochastic optimization (2017),
  \url{https://arxiv.org/abs/1412.6980}

\bibitem{cosine}
Loshchilov, I., Hutter, F.: {SGDR}: Stochastic gradient descent with warm
  restarts (2017), \url{https://arxiv.org/abs/1608.03983}

\bibitem{mixedprecision}
Micikevicius, P., Narang, S., Alben, J., Diamos, G., Elsen, E., Garcia, D.,
  Ginsburg, B., Houston, M., Kuchaiev, O., Venkatesh, G., Wu, H.: Mixed
  precision training (2018), \url{https://arxiv.org/abs/1710.03740}

\bibitem{unet}
Ronneberger, O., Fischer, P., Brox, T.: {U-Net}: Convolutional networks for
  biomedical image segmentation (2015), \url{https://arxiv.org/abs/1505.04597}

\bibitem{covshift2}
Shimodaira, H.: Improving predictive inference under covariate shift by
  weighting the log-likelihood function. Journal of Statistical Planning and
  Inference  \textbf{90}(2),  227–244 (Oct 2000).
  \doi{10.1016/s0378-3758(00)00115-4},
  \url{http://dx.doi.org/10.1016/S0378-3758(00)00115-4}

\bibitem{covshift}
Sugiyama, M., Krauledat, M., M{{\"u}}ller, K.R.: Covariate shift adaptation by
  importance weighted cross validation. Journal of Machine Learning Research
  \textbf{8}(35),  985--1005 (2007),
  \url{http://jmlr.org/papers/v8/sugiyama07a.html}

\bibitem{Thompson2018}
Thompson, M.K., Poortmans, P., Chalmers, A.J., Faivre-Finn, C., Hall, E.,
  Huddart, R.A., Lievens, Y., Sebag-Montefiore, D., Coles, C.E.:
  Practice-changing radiation therapy trials for the treatment of cancer: where
  are we 150 years after the birth of {M}arie {C}urie? British Journal of
  Cancer  \textbf{119}(4),  389--407 (Aug 2018).
  \doi{10.1038/s41416-018-0201-z},
  \url{https://doi.org/10.1038/s41416-018-0201-z}

\bibitem{Traverso2018}
Traverso, A., Wee, L., Dekker, A., Gillies, R.: Repeatability and
  reproducibility of radiomic features: A systematic review. International
  Journal of Radiation Oncology, Biology, Physics  \textbf{102}(4),
  1143–1158 (Nov 2018). \doi{10.1016/j.ijrobp.2018.05.053},
  \url{http://dx.doi.org/10.1016/j.ijrobp.2018.05.053}

\bibitem{dataset}
Wahid, K., Dede, C., Naser, M., Fuller, C.: Training dataset for {HNTSMRG} 2024
  challenge (2024). \doi{10.5281/ZENODO.11199559},
  \url{https://zenodo.org/doi/10.5281/zenodo.11199559}

\bibitem{Wahid2021}
Wahid, K.A., He, R., McDonald, B.A., Anderson, B.M., Salzillo, T., Mulder, S.,
  Wang, J., Sharafi, C.S., McCoy, L.A., Naser, M.A., Ahmed, S., Sanders, K.L.,
  Mohamed, A.S., Ding, Y., Wang, J., Hutcheson, K., Lai, S.Y., Fuller, C.D.,
  van Dijk, L.V.: Intensity standardization methods in magnetic resonance
  imaging of head and neck cancer. Physics and Imaging in Radiation Oncology
  \textbf{20},  88–93 (Oct 2021). \doi{10.1016/j.phro.2021.11.001},
  \url{http://dx.doi.org/10.1016/j.phro.2021.11.001}

\bibitem{edgeart}
Wang, S., Liu, X., Li, Y., Sun, X., Li, Q., She, Y., Xu, Y., Huang, X., Lin,
  R., Kang, D., Wang, X., Tu, H., Liu, W., Huang, F., Chen, J.: A deep
  learning-based stripe self-correction method for stitched microscopic images.
  Nature Communications  \textbf{14}(1) (Sep 2023).
  \doi{10.1038/s41467-023-41165-1},
  \url{http://dx.doi.org/10.1038/s41467-023-41165-1}

\bibitem{staple}
Warfield, S., Zou, K., Wells, W.: Simultaneous truth and performance level
  estimation ({STAPLE}): An algorithm for the validation of image segmentation.
  IEEE Transactions on Medical Imaging  \textbf{23}(7),  903–921 (Jul 2004).
  \doi{10.1109/tmi.2004.828354},
  \url{http://dx.doi.org/10.1109/TMI.2004.828354}

\end{thebibliography}

\end{document}